\documentclass[5p,article]{elsarticle}

\usepackage{graphicx}
\usepackage{mathtools}
\usepackage{verbatim}
\usepackage{amsmath}
\usepackage{caption}
\usepackage{subcaption}
\usepackage[title]{appendix}
\usepackage{physics}
\usepackage[compat=1.0.0]{tikz-feynman}
\usepackage{float}
\usepackage{amssymb}
\usepackage{tabularx}
\usepackage{multirow}
\usepackage{hyperref}
\hypersetup{%
  colorlinks=true, 
  breaklinks=true, 
  citecolor=blue, 
} 

\begin{document}

\title{Sensitivity improvement in hidden photon detection using resonant cavities}
\author[1,2]{Younggeun Kim}
\author[1]{SungWoo Youn\corref{ca}}\cortext[ca]{Corresponding author}\ead{swyoun@ibs.re.kr}
\author[1,2]{Danho Ahn}
\author[1,2]{Junu Jung}
\author[1,2]{Dongok Kim}
\author[1,2]{Yannis K. Semertzidis}

\address[1]{Center for Axion and Precision Physics Research, IBS, 193 Munji-Ro, Daejeon, South Korea, 34051}
\address[2]{Department of Physics, KAIST, 291 Daehak-Ro, Daejeon, South Korea, 34141}

\begin{abstract}
Analogous to the light-shining-through-wall setup proposed for axion-like particle searches, a pair of resonant cavities have been considered to search for an extra U(1) massive gauge boson, called a hidden photon, which mediates the interactions in the hidden sector.
We propose a new cavity configuration, consisting of a cylindrical emitter surrounded by a hollow cylindrical detector to remarkably improve experimental sensitivity to hidden photon signals in the $\mu$eV mass range.
An extensive study was conducted to find the optimal cavity geometry and resonant mode, which yields the best performance.
In addition, a feasible application of superconducting RF technology was explored.
We found the integration of these potential improvements will enhance the sensitivity to the effective kinetic mixing parameter between the hidden photon and the Standard Model photon by multiple orders of magnitude.
\end{abstract}
\maketitle

\section{Introduction}
In the 1980s, a new species of photon was proposed to explain the spectral shape of the cosmic background radiation by extending the Standard Model (SM) with an additional U(1) gauge symmetry~\cite{Georgi:1983aa}.
The corresponding spin-1 boson, called the hidden photon (HP), is assumed to act as a force mediator among particles in the hidden sector (HS), hence the name.
Even if the strength is feeble, the coupling of this gauge boson to the electric charge of the SM particles via kinetic mixing with the SM counterpart, ordinary photons, could provide a portal to the invisible hidden sector~\cite{1982ZhETF..83..892O, HOLDOM1986196}.
For instance, if it interacts with particles in the dark sector, it would shed light on the nature of dark matter, or, if lightly massive, the dark photon itself could account for a significant fraction of cold dark matter.
In addition, the tantalizing tension between theory and experiment regarding the anomalous magnetic moment of the muon has drawn attentions to the possible existence of this gauge particle~\cite{PhysRevD.99.115001}.
A variety of search activities have been conducted considering mass and experimental  signatures.
Searches for heavy mass HPs in the GeV scale have utilized particle colliders~\cite{PhysRevLett.119.131804, PhysRevLett.120.061801}, beam dumps~\cite{PhysRevLett.123.121801}, and fixed targets~\cite{PhysRevD.80.075018, PhysRevD.80.095024}.
On the other hand, light HPs typically in the sub-eV range are sought at smaller-scale labs employing the photon regeneration scheme~\cite{2013PhRvD..88g5014B, PhysRevLett.105.171801, PhysRevD.82.052003}.

In general, the Lagrangian describing the unified framework of the visible and hidden sectors at low energies is given by
\begin{equation}
\begin{split}
\mathcal{L} =& -\frac{1}{4}F_{\mu\nu}F^{\mu\nu} - \frac{1}{4}F'_{\mu\nu}F'^{\mu\nu} - \frac{1}{2}\chi F_{\mu\nu}F'^{\mu\nu}\\
&+ \frac{1}{2}m_{A'}^{2}A'_{\mu}A'^{\mu} - eJ_{\mu}^{em}A_{\mu},
\end{split}
\label{eq:lagrangian}
\end{equation}
where $F_{\mu\nu}$ ($F'_{\mu\nu}$) is the field strength tensor of the SM (HS) gauge field $A_{\mu} (A'_{\mu})$, $m_{A'}$ is the the mass of the HS gauge boson, and $J_{\mu}^{em}$ is the electromagnetic (EM) charged current that $A_{\mu}$ couples to with coupling strength $e$.
The kinetic mixing parameter $\chi$ in the third term is responsible for effective mixing between the ordinary and hidden gauge fields, leading to the interactions between the visible and invisible sectors.
To the first-order approximation of $\chi$, Eq.~\ref{eq:lagrangian} can be reformulated based on the mass eigenstate basis by linear transformations, $A_{\mu}\rightarrow A_{\mu}-\chi A'_{\mu}$ and $A'_{\mu}\rightarrow {A'}_{\mu}$. 
On this basis, the kinetic term and mass term of these fields are diagonalized and the Lagrangian becomes
\begin{equation*}
\mathcal{L}=-\frac{1}{4}F^{2}-\frac{1}{4}F'^{2}+\frac{1}{2}m_{A'}^{2}A'^{2}-eJ_{\mu}^{em}(A_{\mu}-\chi A'_{\mu}).
\end{equation*}
The last term implies that the SM photon ($\gamma$) can be converted to the HS photon ($\gamma'$) or vice versa via the current of charged fermions. 

The HP as dark matter in our Galactic halo can be detected using the conventional axion haloscope technique at no cost to a strong magnetic field.
However, unlike the axion, which is a pseudo-scalar, the vector property of the gauge particle makes a dependence of detection sensitivity on orientation of the cavity resonant mode with respect to the unknown spin polarization.
On the other hand, as a mediator of interactions between HS particles, HPs can be generated in the laboratory and thus the spin polarization can be under control (at least known) depending on how they are generated.
Inspired by the concept of photon regeneration for axion-like particles (ALPs), e.g., “light shining through wall” (LSW) experiments, a detection scheme to search for HPs was proposed that employed two electromagnetically isolated conducting cavities tuned to the same frequency~\cite{1987PhRvL..59..759V, JAECKEL2008509}.
Hidden photons converted from high-density EM photons in one cavity resonator by aid of the surface current propagate in space until reaching the other resonator where they convert back into EM photons.
With $\gamma'$ serving as a mediator, this schematic configuration forms a $\gamma$-$\gamma'$-$\gamma$ oscillation induced by charged currents, analogous to the $\gamma$-$a$-$\gamma$ oscillation induced by strong magnetic fields.
We propose a new cavity configuration based on the fact that the generated HPs propagate isotropically in space.

LSW-type HP search experiments typically import a high-power RF signal generator to supply EM photons to one cavity (emission cavity) and an RF antenna to receive the reconverted EM photon signal in the other (detection cavity)~\cite{PhysRevLett.105.171801}.
The expected oscillation power can be generally expressed as
\begin{equation}
P_{\gamma-\gamma'-\gamma}=\chi^{4}Q_dQ_e (1-\rho^2)^4|\mathcal{G}|^{2}P_{\rm in},
\label{eq:power}
\end{equation}
where $Q_{d,\ e}$ is the cavity quality factor with subscripts $d$ and $e$ denoting the detector and emitter, and $P_{\rm in}$ is the RF power injected into the emission cavity.
The dimensionless parameter $\rho\equiv k_{\gamma'}/k_{\gamma}$, with $k=2\pi/\lambda$ denoting the wave number of a photon of wavelength $\lambda$, is bounded between 0 and 1 to satisfy the condition $k_{\gamma}^{2}=k_{\gamma'}^{2}+m_{\gamma'}^{2}$ on the mass shell.
The factor $\mathcal{G}$ encodes the geometric configuration of the pair of cavities:
\begin{equation}
\mathcal{G}=k_{\gamma}^{2}\int_{V_{d}}\int_{V_{e}}d^{3}\vec{x}_{d}d^{3}\vec{x}_{e}\frac{e^{ik_{\gamma}\rho\abs{\vec{x}_{d}-\vec{x}_{e}}}}{4\pi\abs{\vec{x}_{d}-\vec{x}_{e}}}\vec{A}_{d}(\vec{x}_{d})\cdot\vec{A}_{e}(\vec{x}_{e}),
\label{eq:geo_factor}
\end{equation}
where $\vec{A}$ is the normalized eigen-vector of the vector potential corresponding to the cavity resonant mode. 

Similar to cavity-based axion experiments, the signal-to-noise ratio (SNR) provides a measure of experimental sensitivity:
\begin{equation}
{\rm SNR} = \frac{P_{\rm sig}}{\delta P_{\rm sys}},
\label{eq:snr}
\end{equation}
where $P_{\rm sig}$ $(P_{\rm sys})$ is the detected signal (system noise) power. 
The system noise, $\delta P_{\rm sys}$, is defined as fluctuations in system noise power obtained within integration time $\tau$ over frequency bandwidth $b$ and dictated by the Dicke radiometer equation~\cite{doi:10.1063/1.1770483}:
\begin{equation}
\delta P_{\rm sys}=k_BT_{\rm sys}\sqrt{\frac{b}{\tau}},
\label{eq:noise}
\end{equation}
using the Boltzmann constant $k_B$ and the system noise temperature $T_{\rm sys}$.
The major sources of system noise are the thermal noise of the cavity and the shot noise of the receiver chain. The former is attributed to black-body radiation due to the cavity’s  physical temperature and represented by the effective noise temperature $T_{\rm eff}$, which includes the zero-point quantum fluctuations, as
\begin{equation*}
k_BT_{\rm eff}=h\nu\left(\frac{1}{e^{h\nu/k_BT}-1}+\frac{1}{2}\right),
\end{equation*}
where $h$ is the Plank constant and $\nu$ is the photon frequency.
The latter accounts for the noise added by the RF readout chain, particularly the first stage amplifier, and is represented by the equivalent noise temperature $T_{\rm add}$.
These two noise components linearly contribute to the total system noise, i.e., $T_{\rm sys} = T_{\rm eff} + T_{\rm add}$ \cite{2020JCAP...03..066K}.

An in-depth study of the geometric configuration and electromagnetic properties of cavity modes was performed to determine the optimal setup \cite{PhysRevD.82.052003}.
They demonstrated the viability of the LSW-type experiment at room temperature using two cylindrical copper cavities stacked axially with a choice of the TE$_{011}$ resonant mode to excite the cavities.
ADMX reported a HP search result from their axion dark matter search detectors relying on the TM$_{010}$ resonant mode \cite{PhysRevLett.105.171801}.
By bringing the detection cavity to a cryogenic temperature, they improved the experimental sensitivity and placed more stringent limits on kinetic couplings for masses $<3\,\mu$eV.
In this article, we perform extensive studies to find the optimal cavity configurations and then tune the experimental parameters in Eq.~\ref{eq:power} to enhance the sensitivity even further.

\section{Geometric effect}
\label{sec:geometric_effect}
The geometric factor has a strong dependence on cavity configuration and the resonant mode of interest.
The Green's function in Eq.~\ref{eq:geo_factor} implies that the generated HP field will propagate isotropically, and thus it is inferred that a geometric configuration consisting of the detection cavity surrounding the emission cavity would be beneficial.  In addition,   the choice of cavity mode could be important, because relative field alignment between the two  cavities influences the geometry factor,  as manifested by  the inner product of   the vector fields in Eq.~\ref{eq:geo_factor}.

\subsection{Geometric configuration}
\subsubsection{Spherical geometry}
For an isotropically propagating hidden photon field, one could intuitively imagine a spherical resonator surrounded by a co-central hollow spherical shell, such that the entire outgoing flux from the inner cavity can be captured by the outer shell cavity. 
Such a spherical structure still supports two types of transverse modes, TM and TE modes. 
The allowed EM fields from Maxwell's equations can be decomposed into radial and angular components in the spherical coordinate system as~\cite{2013JEAA....5...32Z}
\begin{equation*}
\begin{split}
\vec{E}_{lm}^{\rm TE} &= (\mathcal{A}_l^{\rm TE} j_{l}(kr) + \mathcal{B}_l^{\rm TE} y_{l}(kr)) \vec{X}_{lm}(\theta,\phi)\\
\vec{B}_{lm}^{\rm TM} &= (\mathcal{A}_l^{\rm TM} j_{l}(kr) + \mathcal{B}_l^{\rm TM} y_{1}(kr)) \vec{X}_{lm}(\theta,\phi),
\end{split}
\end{equation*}
where $j_l$ and $y_l$ are the spherical Bessel functions of the first and second kind, respectively, and $\vec{X}_{lm}$ is the vector spherical harmonics. 
The coefficients $\mathcal{A}_l^{\rm TE/TM}$ and $\mathcal{B}_l^{\rm TE/TM}$ are determined by the boundary conditions of the geometry.
For convenience sake, the cavity thickness and the gap between the cavities are ignored.
The lowest TE and TM modes correspond to $(l, m)=(1,0)$. 
If the inner and outer cavities are tuned at the same resonant frequency, the only solution of the radial component is $j_{1}(kr)$ since $y_{1}(kr)$ is not defined at $r=0$.
The $E$-field of the TE$_{10}$ mode is given to be
\begin{equation*}
\vec{E}_{10}^{\rm TE}=\mathcal{A}_1^{\rm TE} j_{1}(kr)\sin\theta\hat{\phi},
\end{equation*}
with a boundary condition $j_{1}(kR_i)=j_{1}(kR_o)=0$, where $R_i$ and $R_o$ are the radii of the inner and outer cavities.
The $E$-field of the TM$_{10}$ mode, on the other hand, is obtained by Ampere's law  $\nabla\times\vec{B}^{\rm TM}_{lm}=ik\vec{E}_{lm}^{\rm TM}$:
\begin{equation*}
\vec{E}_{10}^{\rm TM}=\mathcal{B}_1^{\rm TM} \left(\frac{2\cos\theta}{kr}j_{1}(kr)\hat{r}-\frac{\sin\theta}{kr}\left(\sin kr-j_{1}(kr)\right)\hat{\theta} \right), 
\end{equation*}
where two boundary conditions, $\sin (kR_i)=j_{1}(kR_i)$ and $\sin (kR_o)=j_{1}(kR_o)$, are satisfied.

\begin{figure}
\centering
\includegraphics[width=0.8\linewidth]{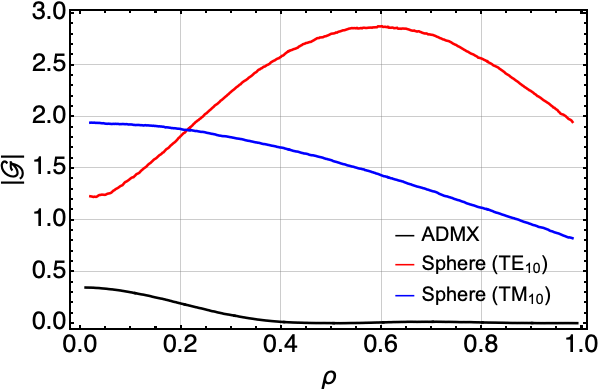}
\caption{$\mathcal{G}$ factors for the two lowest modes of the spherical geometry, TE$_{10}$ and TM$_{10}$, as a function of $\rho$.
That for the TM$_{010}$ with the ADMX configuration is also drawn for comparison.}
\label{fig:G_sphere}
\end{figure}

A Coulomb gauge with no source present, i.e., $A_{0}=0$, relates the electric field with the vector potential in a simple manner as
\begin{equation*}
\vec{E}=-\frac{\partial\vec{A}}{\partial t}.
\end{equation*}
Using the Mathematica software package~\cite{mathematica}, the $\mathcal{G}$ factors of the lowest TE and TM modes for the spherical geometry are calculated as a function of the relative momentum $\rho$ and compared with that of the ADMX geometry in Fig.~\ref{fig:G_sphere}. 
The lower and upper bounds, $0\le \rho \le 1$, correspond to the most massive ($m_{\gamma'}=k_\gamma$) and massless ($m_{\gamma'}=0$) HPs, respectively.
It is noted that for a chosen resonant mode of a spherical geometry, the dimension of the detector is determined in subordination to that of the emitter and thus the $\mathcal{G}$ factor is independent of their ratio, $R_o/R_i$.

\begin{figure}
\centering
\includegraphics[width=0.7\linewidth]{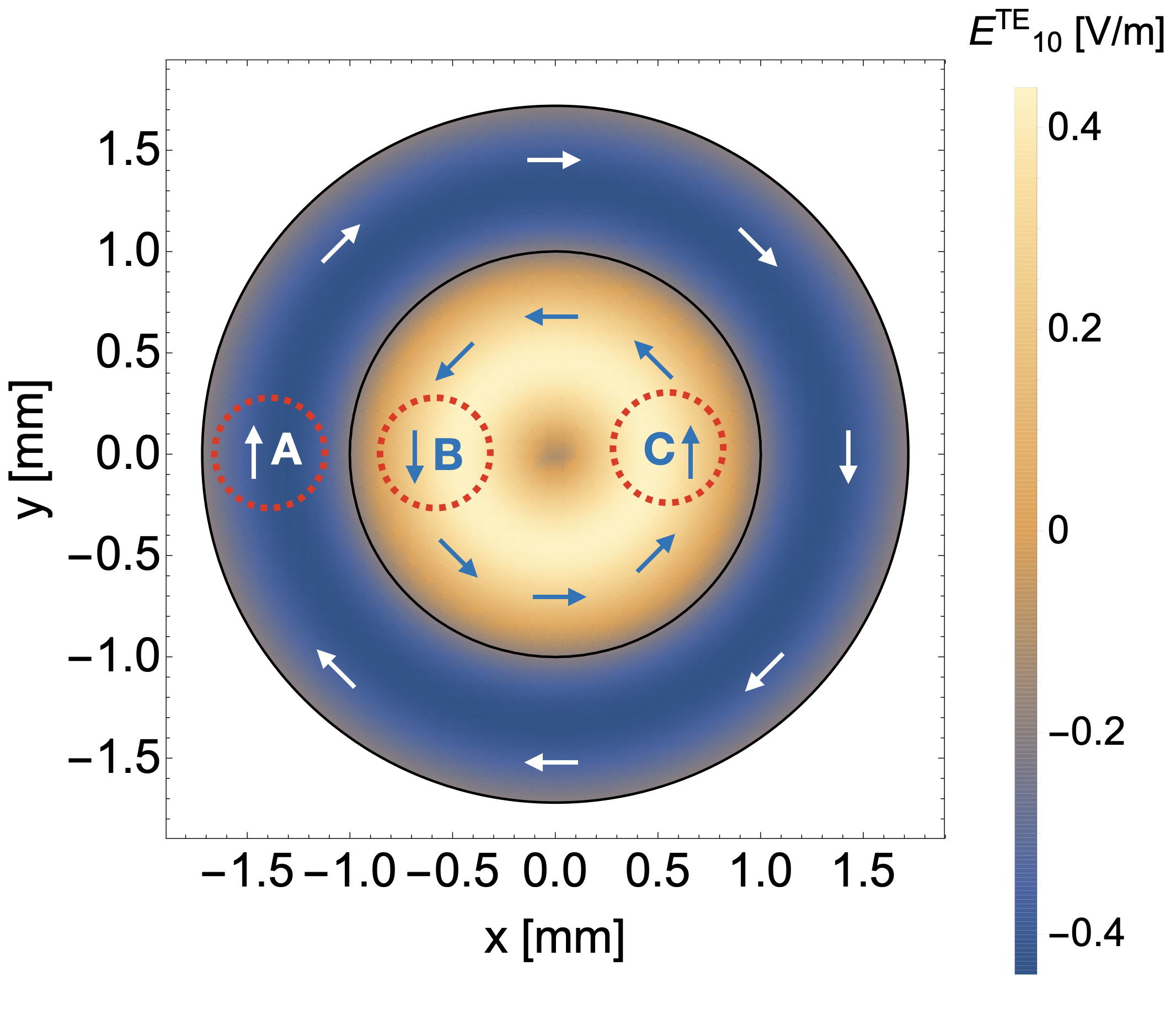}
\caption{Cross-sectional view of the spherical geometry with the $E$-field distributions corresponding to the TE$_{10}$ mode.
The field directions are represented by the arrows.
The field component marked as `A' in the outer cavity oscillates in the opposite (same) direction to (as) the one marked as `B' (`C') in the inner cavity.
}
\label{fig:field_sphere}
\end{figure}

In this configuration, however, even though the outer cavity covers the total flux from the inner cavity, there are always field components in one cavity which oscillate in different directions with respect to a field component in the other, as illustrated in Fig.~\ref{fig:field_sphere}.
These combinations produce cancellation effects in the overall integration of the inner product in Eq.~\ref{eq:geo_factor} making $\mathcal{G}$ not maximal. 
Furthermore, a tuning mechanism to synchronize both cavities at the same frequencies would not be trivial.
These issues lead us to consider a cylindrical geometry which supports well-defined transverse modes and facilitates cavity alignment for field arrangement in a desired way.
In addition, the tuning mechanisms could be implemented in a more straightforward way.

\subsubsection{Cylindrical geometry}
\begin{figure}
\centering
\begin{subfigure}{0.4\linewidth}
\includegraphics[width=\linewidth]{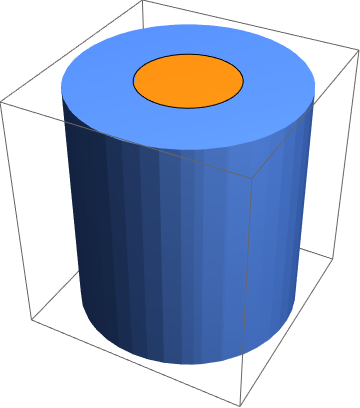}
\caption{}
\label{fig:geometry_cylinder_1}
\end{subfigure}
\begin{subfigure}{0.55\linewidth}
\includegraphics[width=\textwidth]{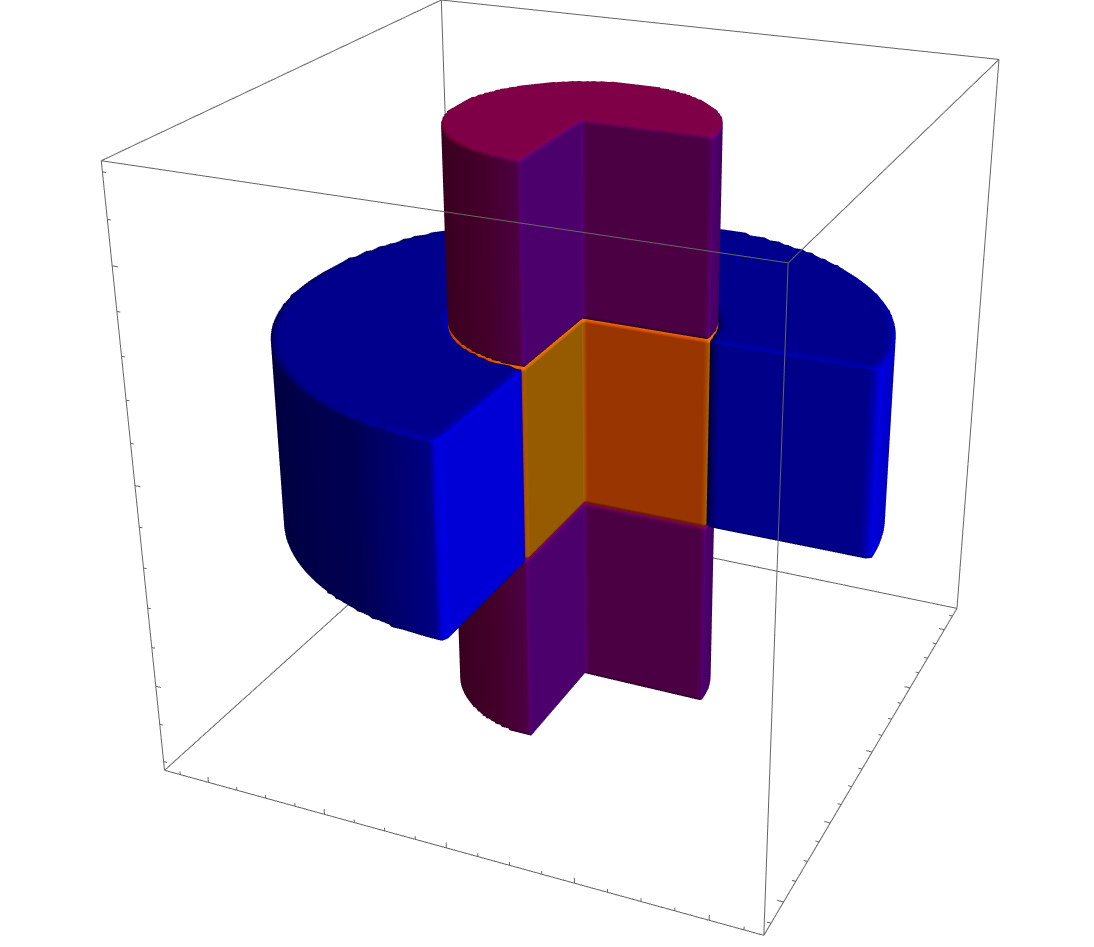}
\caption{}
\label{fig:geometry_cylinder_2}
\end{subfigure}
\caption{Two-cavity configurations using a cylindrical geometry for hidden photon searches.   (a)   A hollow cylinder (blue) surrounds an inner cylinder (orange).   (b) The inner cylinder is encapsulated by the hollow cylinder with additional cylinders (purple) on the top and bottom.}
\label{fig:geometry_cylinder}
\end{figure}

In a cylindrical resonator, the EM fields of some resonant modes are defined in a particular direction in the Coulomb gauge.
For example, the electric (magnetic) field of the TM$_{0n0}$ (TE$_{0nm}$) modes only has the $z$ component.
Therefore, to maximize the $\mathcal{G}$ factor, it would be easier to configure a two-cavity system so that the vector fields of the individual cavities oscillate in the same direction.
The stacked configuration considered in Ref.~\cite{PhysRevD.82.052003}, employed the TE$_{011}$ mode to take advantage of the effect.
However, this geometry was not optimal in terms of acceptance of the HP flux radiating out from the emitter.
Here, we consider a new cavity configuration where a cylinder for emission is surrounded by a hollow cylinder for detection, as shown in Fig.~\ref{fig:geometry_cylinder_1}.
The dimensions of the hollow detection cylinder are determined relatively to those of the emission cylinder, so that they have the same frequency of the resonant mode under consideration.
An improved geometric configuration is also conceivable by introducing an additional pair of cylinders stacked on the top and bottom of the cylinder to completely capsulate the emitter (see Fig.~\ref{fig:geometry_cylinder_2}). Combined with the hollow cylinder this would result in a three-piece detector.
Assuming the quality factors of the three individual detection cavities are the same, the total conversion power becomes the linear sum of the individual contributions:
\begin{equation*}
P_{\rm{sig}} = P_h + 2P_s \propto \abs{\mathcal{G}_h^2 + 2\abs{\mathcal{G}_s}}^2,
\end{equation*}
where the subscripts $h$ and $s$ denote the hollow and stacked cylinders, respectively.
The simplest design requires the height of both the stacked and hollow cylinders to be the same as the emission cavity.
Again, the cavity thickness and the gap between the cavities are ignored for simplicity. 

The general field solutions for the transverse modes of a cylinder with radius $R$ and length $L$ are given as
\begin{equation*}
\begin{split}
&\vec{E}_{lnm}^{\mathrm{TM}}=\psi(r,\phi)\cos(\frac{m\pi z}{L})\hat{z}-\frac{m\pi}{L\alpha^{2}}\sin(\frac{m\pi z}{L})\vec{\nabla}_{t}\psi(r,\phi) \\
&\vec{E}_{lnm}^{\mathrm{TE}}=-\frac{i\omega\mu}{\alpha^{2}}\sin(\frac{m\pi z}{L})\hat{z}\times\vec{\nabla}_{t}\psi(r,\phi),
\end{split}
\end{equation*}
The scalar function $\psi(r,\phi)$ is expressed in terms of the Bessel functions of the first ($J_l$) and second kind ($Y_l$) as
\begin{equation*}
\psi(r,\phi) = \left[\mathcal{A}_{ln} J_{l}(k_{n}r) + \mathcal{B}_{ln} Y_{l}(k_{n}r)\right] e^{\pm il\phi}
\end{equation*}
with respective coefficients $\mathcal{A}_{ln}$ and $\mathcal{B}_{ln}$.
The constant $\alpha$ is defined as
\begin{equation*}
\alpha^{2} \equiv \mu\epsilon\omega^2 - \left(\frac{m\pi}{L}\right)^{2},
\end{equation*}
where the permeability $\mu$ and permittivity $\epsilon$ of the system and the angular frequency $\omega(=2\pi\nu)$ of the EM wave.
The wave function $\psi$ satisfies the two-dimensional differential equation $(\nabla_{t}^{2}+\alpha^{2})\psi(r,\phi)=0$, where $\vec{\nabla}_{t}$ denotes the gradient operator in the transverse plane perpendicular to the $z$ direction.
For the cylindrical cavity, the function $\psi$ can have only the $J_l$ component since $Y_l$ is ill-defined at $r=0$.
In particular, for $(l,m) = (0,0)$, the $E$-field of the TM mode simply becomes
\begin{equation*}
\vec{E}_{0n0}^{\mathrm{TM}}(r,\phi,z) = \mathcal{A}_{0n}^{\rm TM} J_{0}(k_{n}r)\hat{z},
\end{equation*}
which requires $k_{n}=\chi_{0n}/R$ where $\chi_{0n}$ is the $n$-th root of the Bessel function $J_{0}(k_{n}r)$.
The TE$_{0nm}$ modes, on the other hand, have an azimuthally oscillating field component:
\begin{equation*}
\vec{E}^{\mathrm{TE}}_{0nm}(r,\phi,z)=\mathcal{A}_{0n}^{\rm TE}k_{n}J_0(k_{n}r)\sin(\frac{m\pi z}{L})\hat{\phi},
\end{equation*}
with the boundary condition of $J_0(k_nR)=0$.

\begin{figure}
\centering
\includegraphics[width=0.8\linewidth]{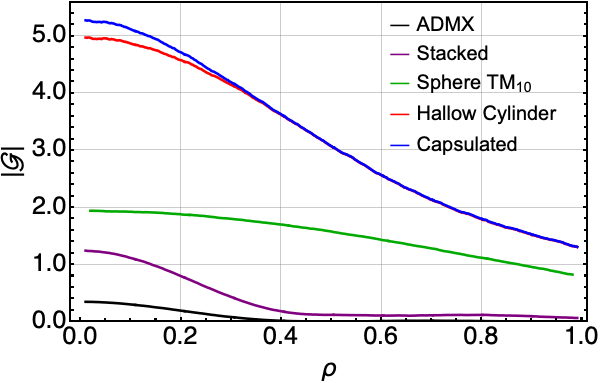}

\vspace{0.025\linewidth}

\includegraphics[width=0.8\linewidth]{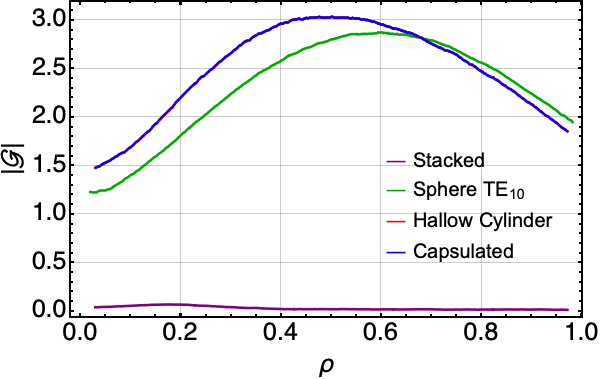}
\caption{Comparison of the $\mathcal{G}$ factor as a function of $\rho$ for several cavity configurations using the dimensions described in Table~\ref{tab:dim_geo}.
The two most general resonant modes are considered: TM$_{010}$ (TM$_{10}$) (top) and TE$_{011}$ (TE$_{10}$) (bottom) for the cylindrical (spherical) geometry. 
For the bottom figure, the $\mathcal{G}$ factors for ``Hollow cylinder" and ``Capsulated" are overlapped.
}
\label{fig:G_comp}
\end{figure}

The performances of various detector configurations were compared for a given emission cavity which has the same dimensions as the ADMX emitter. 
The dimensions of the individual detection cavities are summarized in Table~\ref{tab:dim_geo}.
The $\mathcal{G}$ factors are computed using Mathematica for the two mostly considered lowest modes, TM$_{010}$ and TE$_{011}$. They are compared in Fig.~\ref{fig:G_comp}, where it is found that the hollow cylindrical cavity remarkably improved the performance over the entire range.
The spherical geometry yielded higher values than either the ADMX or stacked configurations, but not as much as the hollow cylindrical geometry, as expected.
The capsulated version, meanwhile, did not produce a noticeable improvement compared to that with the hollow cylinder, only due to the non-significant contributions from the stacked cavities, as manifested in the same figures.
Since the non-capsulated cylindrical geometry can be realized with a simpler structure in a less space, it was chosen as our detector configuration for further studies.

\begin{table}
\begin{center}
\begin{tabular*}{\columnwidth}{@{\extracolsep{\fill}}c@{}c@{}c@{}c}
 \hline
 Geometry & Configuration &Radius [cm] & Length [cm] \\
 \hline
 \hline
 \multirow{4}{*}{Cylinder} & ADMX & 20.0 & 92.6 \\
  & Stacked & 20.0 & 92.6 \\
  & Hollow ($R_o$)& 45.9 & 92.6 \\
  & Capsulated & 20.0\,:\,45.9 & 92.6\,:\,92.6\,:\,92.6 \\
 \multirow{2}{*}{Sphere} & TM ($R_i$\,:\,$R_o$) & 22.8\,:\,50.9 & -\\
  & TE ($R_i$\,:\,$R_o$) & 37.4\,:\,64.2 & - \\
 \hline
\end{tabular*}
\end{center}
\caption{: Dimensions of the detection cavity for various geometric configurations for a given emission cavity whose dimensions are normalized to the ADMX cavity, i.e., 20.0\,cm in radius and 92.6\,cm in length. The capsulated detector is a combined version of the hollow cylinder with a pair of stacked cylinders on the top and bottom of the emitter. For a spherical geometry, they are determined to have the same resonant frequencies as those of a cylindrical geometry.}
\label{tab:dim_geo}
\end{table}

\subsection{Aspect ratio and resonant mode}
The effect of geometry on detection power is attributed to not only the shape and configuration of the two cavities, but also the relative dimensions of the individual cavities.
The relevant quantity is the aspect ratio and for the cylindrical geometry it is given by $\beta = 2R/L$.
In general, the quality factor and the resonant frequency of a cavity also vary with the ratio.
In particular, the resonant frequency is associated with the photon momentum, which influences the HP momentum $\rho$ for a given HP mass.
Therefore, from Eq.~\ref{eq:power}, we can define a more general $\beta$-dependent geometry factor
\begin{equation}
\mathcal{F}(\beta)\equiv Q_d(\beta)Q_e(\beta)[1-\rho(\beta)^2]^4\abs{\mathcal{G}(\beta)}^{2},
\label{eq:f_factor}
\end{equation}
as a figure of merit for the study.
Since, as indicated in Figs.~\ref{fig:G_sphere} and~\ref{fig:G_comp}, the $\mathcal{G}$ factor is also sensitive to the choice of the cavity mode, we examined the $\beta$ effect for several modes: two reference modes, TM$_{010}$ and TE$_{011}$, and some of their higher modes, TM$_{020}$, TM$_{030}$ and TE$_{021}$.

\begin{figure}
\centering
\includegraphics[width=0.8\linewidth]{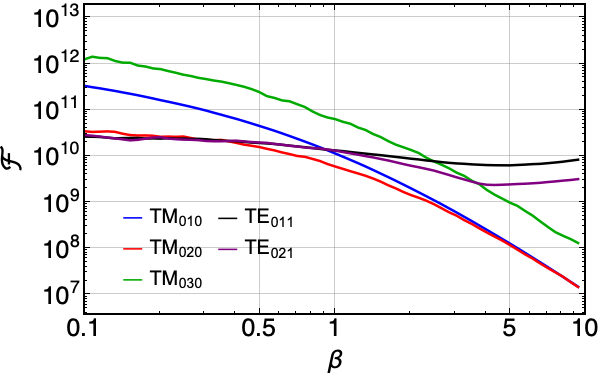}
\caption{Comparison of the $\mathcal{F}$ factor for several resonant modes of the hollow cylindrical geometry as a function of $\beta$.}
\label{fig:aspect_ratio}
\end{figure}

Considering the hollow cylinder geometry, we altered the height of the emission cavity for a fixed radius to vary the aspect ratio.
The height of the detection cavity was also altered accordingly.
Assuming the cavities are made of copper, the $\mathcal{F}$ factors were evaluated for different resonant modes as a function of $\beta$, and are compared in Fig.~\ref{fig:aspect_ratio}. 
The TM modes generally show in general a strong dependence on $\beta$ with high performance at small values, while the TE modes are less sensitive to $\beta$ and will be beneficial for large values.
Interestingly, a somewhat unexpected behavior appears for the higher-order modes.
Since the higher-degree variations in the EM fields can create destructive contributions which suppress the geometry factor, one might expect that the higher modes would yield lower performance than the fundamental mode, just like axion haloscope searches where the form factor is suppressed for higher-order modes because of negatively oscillating field components under a uniform magnetic field.
However, we observed that the TM$_{030}$ mode yielded higher performance than the TM$_{010}$ mode, which is the opposite of the case observed with axion haloscopes where the form factor is even more suppressed than the TM$_{020}$ mode~\cite{Kim_2020}.

\begin{figure}[b]
\centering
\includegraphics[width=0.8\linewidth]{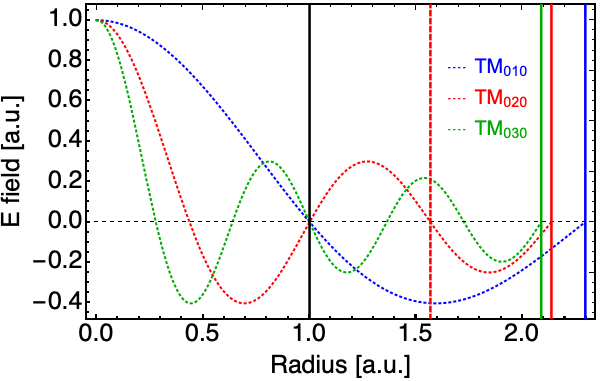}
\caption{$E$-field profiles (colored dotted lines) of the TM$_{0n0}$ mode for the cylindrical geometry. 
The black vertical line at $R_{\rm rel}=1.0$ indicates the common boundary between the emitter and the detectors.
There are multiple choices for the outer radius of the detector -- the radius bound by the colored vertical solid (dashed) lines yield the EM solutions with $n$th-order (1st-order) field variation.
}
\label{fig:TM030_node}
\end{figure}

It is noted that for LSW-type searches, the dot product in Eq.~\ref{eq:geo_factor} is made between two oscillating vector fields, while for axion haloscopes it is made between an oscillating and a static field.
The former is less intuitive than the latter in terms of the field geometry effect.
The net effect of the dot product of the electric fields between the emitter and the detector can be visualized in Fig.~\ref{fig:TM030_node}.
When there is an even number of nodes in field distribution, such as TM$_{020}$, there is an identical number of field combinations to make positive and negative contributions to the overall dot product.
For higher modes with an odd number of node, such as TM$_{030}$, on the other hand, there are more negative combinations, which yield a larger value of $\mathcal{F}$ factor when squared.
Such behaviors were confirmed by examining even higher-order modes up to the TM$_{050}$ mode.

\begin{figure}
\centering
\includegraphics[width=0.8\linewidth]{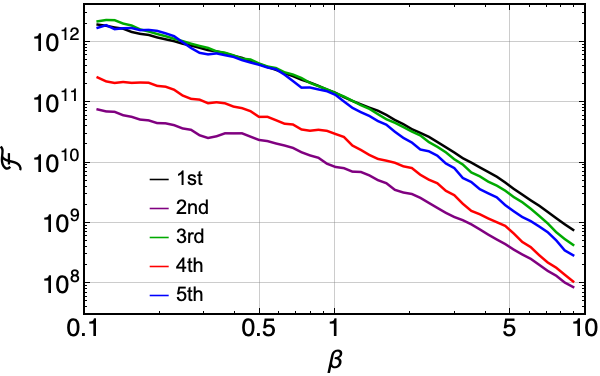}
\caption{$\mathcal{F}$ factors of the TM$_{050}$ mode for different detectors with different outer radii determined in the same way as in Fig.~\ref{fig:TM030_node} depending on the order of field variation.}
\label{fig:TM030-F}
\end{figure}

\begin{figure}[h]
\centering
\includegraphics[width=0.8\linewidth]{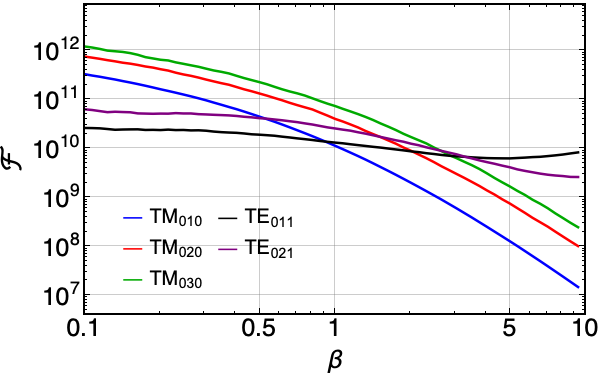}
\caption{Comparison of the $\mathcal{F}$ factor for the same resonant modes in Fig.~\ref{fig:aspect_ratio} with the choice of the detector's outer radius, which satisfies the lowest-order (1st-order) EM field solution.}
\label{fig:aspect_ratio-2}
\end{figure}

In the meantime, we also note that in exploiting higher-order modes, multiple choices are possible when determining the dimension of the detector so that it will have the same resonant frequency as the emitter.
For example, for the TM$_{020}$ mode, there are two  options for  the outer radius depending on the order of field  variation  along  the  radial  direction, as illustrated in Fig.~\ref{fig:TM030_node}.
The larger radius, represented by the vertical solid red line, yields an EM field solution with double anti-nodes (2nd-order variation). This was used to calculate the $\mathcal{F}$ factors in Fig.~\ref{fig:aspect_ratio}. Meanwhile, the smaller detector, bounded by the dashed red line, possesses only a single anti-node (1st-order variation) in its solution.
Since these two solutions give the same resonant frequency, it is not necessary to design the detector so that it has the same order of field variation as the emitter.  The latter choice is expected to induce more unbalanced positive-negative contributions to the overall dot product, as mentioned, resulting in a recovery of the geometry factor.
To verify this in a comprehensive fashion, the TM$_{050}$ mode was chosen to examine how dependent performance was on the choice of the outer radius, which is shown in Fig.~\ref{fig:TM030-F}.
We found the detectors with odd numbers of field variation consistently exhibited higher performance than those with even numbers.
We also noticed that the $\mathcal{F}$ factor with the lowest-order (1st-order) field variation was compatible with those with higher-order (3rd- and 5th-order) field variations.
The $\mathcal{F}$ factors for the higher-order modes in Fig.~\ref{fig:aspect_ratio} were recalculated considering the detector dimension satisfying the lowest-order field solution, and then compared with their lowest modes in Fig.~\ref{fig:aspect_ratio-2}, where considerable improvements were observed.
This indicates that utilizing higher-order resonant modes with this detector configuration will be beneficial for exploring for high-mass HPs, which is a distinctive aspect of cavity experiments.

\section{High-Q resonator}
As indicated by Eq.~\ref{eq:power}, the signal power also depends on the quality factor of the individual cavities and the input power of the RF photons input into the emitter.
The dependency can be addressed in a more straightforward manner than the complicated geometric effect discussed in Sec.~\ref{sec:geometric_effect}. 
For normal conductors, the quality factor is obtained using a relation with the cavity geometry factor $G$ and the surface resistance $R_s$ as
\begin{equation}
Q = \frac{G}{R_s} = \frac{\mu\omega\frac{\int dV |\vec{H}|^{2}}{\oint dS |\vec{H}|^{2}}}{(\sigma\delta)^{-1}},
\label{eq:q_factor}
\end{equation}
where $\mu$ is the permeability inside the cavity, $H$ is the magnetic field strength of the resonant mode.
The surface resistance is determined by the conductivity $\sigma$ and the skin depth $\delta = \sqrt{2 / \mu \sigma \omega}$ in the radio frequency regime. 
In Eq.~\ref{eq:f_factor}, $Q$ was numerically calculated by taking both the cavity geometry and surface resistance into account.

A natural approach to improving the cavity quality factor would be to employ superconducting radio-frequency (SRF) technology.
Various characteristics of SRF cavities, particularly those made of niobium branch superconductors, have been extensively studied~\cite{PhysRevSTAB.3.092001} and successfully described by the BCS theory of superconductivity~\cite{PhysRevLett.113.087001}.
For these superconductors, the surface resistance can be decomposed into two components -- the BCS resistance $R_s^{BCS}$ and the residual resistance $R_s^{res}$.
The former contribution has a strong dependence on temperature and frequency, which is given by 
\begin{equation*}
R_s^{BCS}=4\pi^2\mu^2\nu^2\lambda^3\sigma_{n}\frac{\Delta(T)}{k_BT}\ln\left[\frac{9k_BT}{4h\nu}\right]\exp\left[-\frac{\Delta(T)}{k_BT}\right], 
\end{equation*} 
where $\lambda$ is the London penetration depth, $\sigma_{n}$ is the electrical conductivity in normal state.
The energy gap $\Delta/k_BT$ is a temperature-dependent parameter with respect to the critical temperature $T_c$ and it is known that $\Delta(0)/k_{B}T_{c} = 1.76$ for pure Nb~\cite{PhysRev.128.591}.
The latter contribution to surface resistance, on the other hand, is a residual component attributed to extraneous effects such as material properties and fabrication methods, which can be cured to some degree by heat treatments~\cite{PhysRevSTAB.3.092001, Weingarten:1990bg}.
The residual resistance accounts for losses, which are unrelated to the superconducting state, and is approximately independent of temperature and frequency in the absence of a magnetic field~\cite{PhysRevSTAB.3.092001}.
As a result, the total surface resistance is expressed as $R_s(\nu,T) = R_s^{BCS}(\nu,T)+R_s^{res}$.

A study of RF properties in niobium cavities in the GHz range has reported that, particularly at temperatures $<2$\,K, there is a very slow increment in $R_s$ with surface field up to roughly 100\,mT, beyond which $R_s$ suddenly rises, indicating entrance to the normal state~\cite{doi:10.1063/1.1767295}. 
This implies that the quality factor of an niobium emission cavity will endure the input power of EM waves up to a certain level.
Here we show that the surface field can be analytically determined for a given cavity geometry and injected RF power.
For instance, for the TM$_{010}$ mode of a cylindrical cavity, the quality factor (Eq.~\ref{eq:q_factor}) and resonant frequency are determined respectively by
\begin{equation*}
    Q = \frac{\mu c}{2R_s}\frac{\chi_{01}}{1+R/L}\;\;{\rm and}\;\;\omega=\frac{c\chi_{01}}{R}.
\end{equation*}
If we assume that all the input power is dissipated by loss on the surface, the $Q$-factor is given by definition as 
\begin{equation*}
    Q \equiv \frac{\omega U}{P_{\rm loss}} = \frac{\omega U}{P_{\rm in}},
\end{equation*}
where $U = \mathcal{C} B_s^2$ is the stored energy expressed by the surface magnetic field $B_s$ and the coefficient $\mathcal{C}$, which can be determined by simulation.
Combining the relations above, we derive a formula for $B_s$ as a function of $P_{\rm in}$ as 
\begin{equation*}
    B_s = \left(\frac{\mu P_{\rm in}}{2\mathcal{C}R_s}\frac{RL}{R+L}\right)^{1/2}.
\end{equation*}
For the ADMX cavity geometry, we obtain $\mathcal{C}=3.7\times10^{4}$\,J/T$^2$ from the simulation, which gives $B_s = 0.42\,\mu$T for an input power of 150\,mW assuming $R_s = 6.14\times10^{-3}\,\Omega$ for copper.

\begin{figure}
\centering
\includegraphics[width=0.85\linewidth]{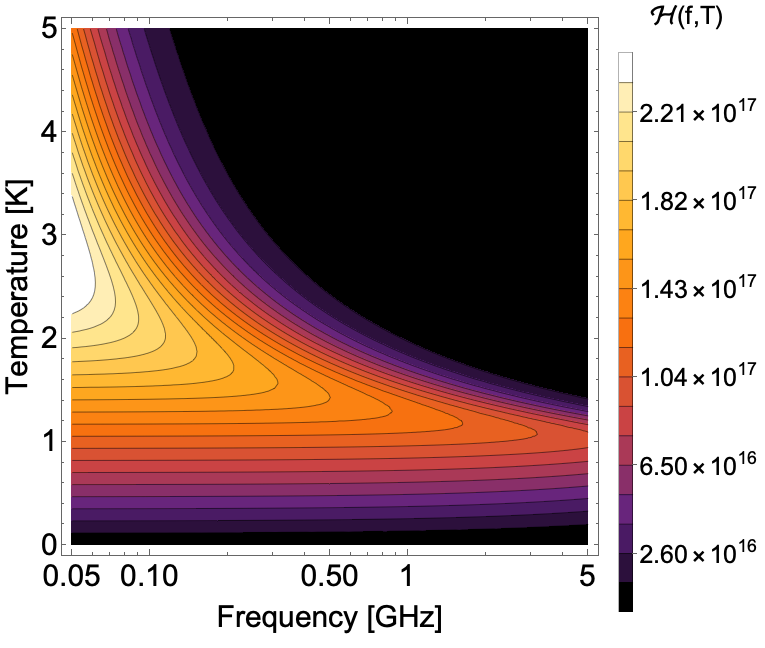}
\label{Full surrounding geometry}
\caption{Computed 2D distribution of $\mathcal{H}(\nu,T)$ for a typical cooling power $P_c=0.5\,{\rm mW}\left(\frac{T}{0.1\,{\rm K}}\right)^2$ under the assumption of $T_{\rm SQL}=h\nu/k_BT$ and $R_s^{res}=3$\,n$\Omega$.}
\label{fig:opt_temp}
\end{figure}

However, the dissipated power on the surface will cause the cavity temperature to rise.
For superconducting cavities, in particular, this may substantially increase the surface resistance and thus the maximum input power will be limited by the cooling capacity of the cryogenic system. 
Now, if we define a function of frequency and temperature, including the noise temperature $T_{\rm sys}$, as
\begin{equation*}
\mathcal{H}(\nu,T)\equiv\frac{P_{\rm in}(T)}{T_{\rm sys}(\nu,T) [R_s^{BCS}(\nu,T)+R_s^{res}]^2},
\end{equation*}
Eq.~\ref{eq:snr} can be reformulated as
\begin{equation*}
{\rm SNR}(\nu,T)=\frac{P_{\rm{sig}}}{k_{B}T_{\rm sys}} \sqrt{\frac{\tau}{b}} = \mathcal{D}\mathcal{H}(\nu,T),
\end{equation*}
with the coefficient $\mathcal{D}$ encompassing all other parameters which are assumed to remain unchanged for a given experimental setup.
For the moment, the surface resistance is also assumed to be independent of the input power.
Since $\mathcal{H}(\nu,T)$ is a numerically calculable function, the optimal $P_{\rm in}(T)$ can be estimated for a given cooling power $P_c$.
For example, commercially available cryogen-free coolers have a typical cooling power of $\sim500\,\mu{\rm W}$ at 100\,mK with a quadratic temperature dependence, i.e.,
\begin{equation}
P_{\rm in, max}(T)\approx P_{\rm c}(T)=500\,\mathrm{\mu W}\left(\frac{T}{0.1\,{\rm K}}\right)^2.
\label{eq:p_max}
\end{equation}
Assuming minimal added noise, represented by a half of the standard quantum limit (SQL), $T_{\rm SQL}=h\nu/k_BT = 48\,{\rm mK}\times \nu/$GHz, and $R_s^{res}=3$\,n$\Omega$ for Nb~\cite{PhysRevSTAB.3.092001}, the function $\mathcal{H}(\nu,T)$ is numerically obtained, as represented by the two-dimensional distribution in Fig.~\ref{fig:opt_temp}.
We notice that the optimal system temperature asymptotically approaches a particular value, $\sim1.1$\,K, at high frequencies, at which Eq.~\ref{eq:p_max} returns the maximum input power to the emitter of $\sim60$\,mW.
This corresponds to a surface field of $2.9\,\rm{m}$T, which falls well within the range where the quality factor is not noticeably degraded.
Finally, a combination of these parameter values will yield the maximum SNR for the given cryogenic system in the GHz range.

\section{Projected sensitivity}
Unlike axion or ALP searches where the particle mass is determined by the searching frequency of the photon in the non-relativistic limit, for HP searches, the particle mass is uncertain within the range $0 < m_{\gamma'} < k_\gamma$ due to its unknown momentum.
Therefore the $\gamma$-$\gamma'$ mixing depends on the HP mass for a given photon frequency.
By plugging Eqs.~\ref{eq:power} and~\ref{eq:noise} into Eq.~\ref{eq:snr}, we express the kinetic mixing parameter $\chi$ as 
\begin{equation}
\chi(\rho) = \left(\frac{k_{B}T_{\rm{sys}}}{Q_{e}Q_{d}P_{\rm{in}}}\sqrt{\frac{b}{\tau}}\frac{\rm SNR}{\abs{\mathcal{G}(\rho)}^{2}}\right)^{1/4} \frac{1}{1-\rho^2},
\label{eq:chi}
\end{equation}
from which we estimate the projected sensitivity as a function of HP mass $m_{\gamma'}=k_{\gamma}\sqrt{1-\rho^{2}}$.
It is found that heavy HPs have higher couplings than light ones.
It is also noted that the mass must be massive, i.e., $\rho<1$, such that the HP becomes distinguishable from the massless SM photon and the coupling constant in Eq.~\ref{eq:chi} does not diverge.

\begin{figure}
\centering
\includegraphics[width=\linewidth]{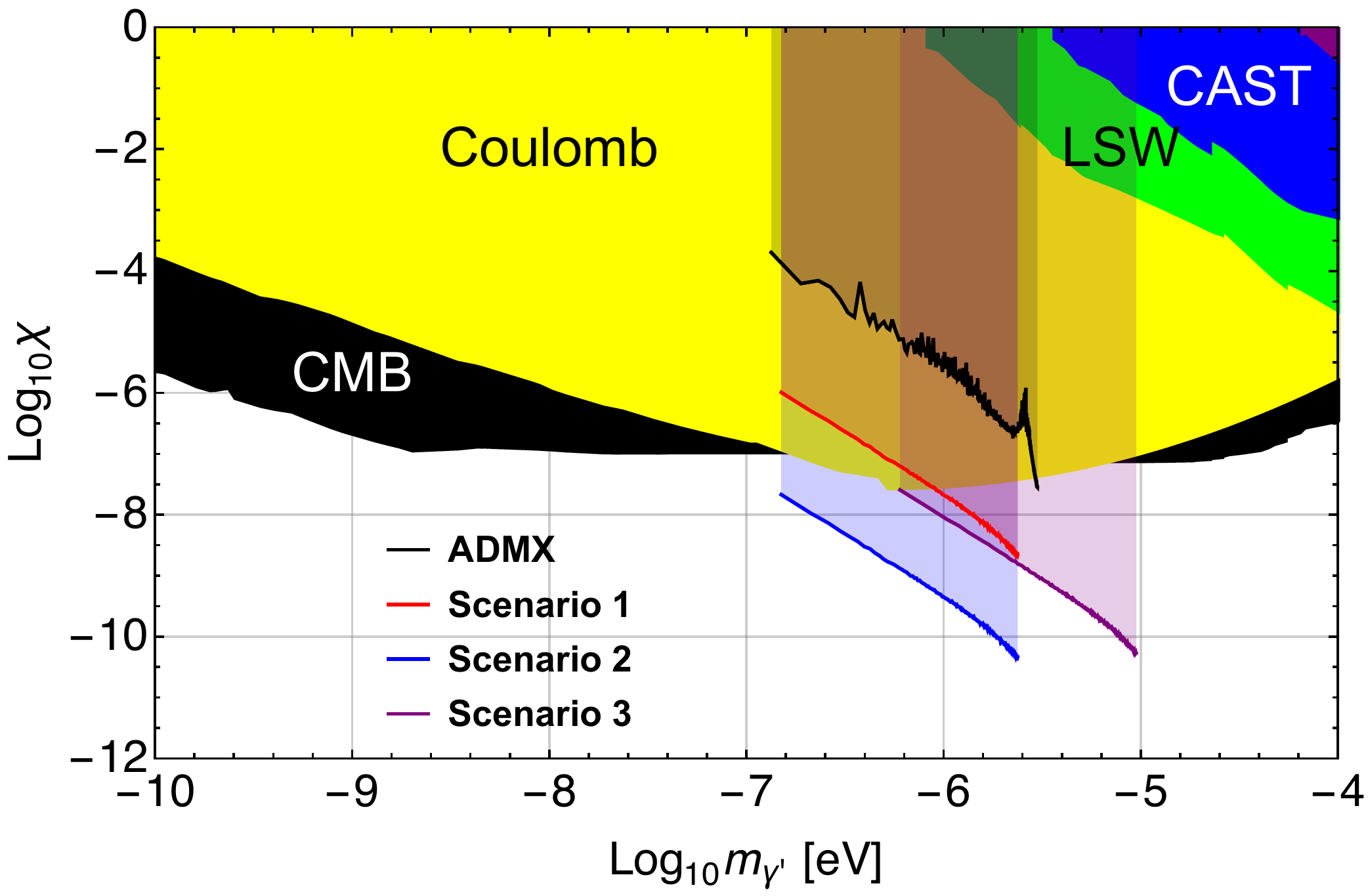}
\caption{
Projected sensitivities for various experimental scenarios, compared with the ADMX result.
Scenario 1 represents the sensitivity improvement using our hollow cylindrical geometry with the same emitter and measurement setups as for ADMX.
Scenario 2 reflects the potential enhancement assuming a quantum-noise limited amplifier and superconducting cavities ($Q=10^8$).
The optimally obtained input power ($P_{\rm in}=60$\,mW) and physical temperature ($T=1.1$\,K) are also implemented. 
Scenario 3 projects an expectation at higher frequencies by scaling the cavity down ($R=5.0$\,cm and $L=23.2$\,cm).
}
\label{fig:sensitivity}
\end{figure}

Adopting our cylindrical geometric configuration, consisting of a cylinder surrounded by a hollow cylinder, the mixing parameter $\chi$ is computed as a function of HP mass, using the same cavity dimensions and measurement parameters as for ADMX, i.e., copper cavities with $P_{\rm{in}} = 150$\,mW, $T_{\rm sys} = 6$\,K, $b= 125$\,Hz, and $\tau = 2$\,h.
After choosing the TM$_{010}$ mode, the expected sensitivities for several scenarios are projected in Fig.~\ref{fig:sensitivity}, where the ADMX result is also compared.
It is remarkable that our geometric configuration alone improves the sensitivity by about two orders of magnitude.
An additional improvement is apparent with the application of quantum science and SRF technology for low noise and high $Q$-factor, respectively.
Upgrading the cryogenic refrigeration systems with higher cooling capacity will also allow for high power injection into the emitter.
It was also observed that the scale of the cavity, which alters the resonant frequency, does not affect the geometry factor as long as the aspect ratio remains unchanged, implying that the experimental sensitivity is weakly dependent on the searching frequency in contrast to axion cavity haloscope experiments.

\section{Conclusion}
We propose a geometric configuration for LSW-type experiments which employs a pair of cavities to search for a new gauge boson in the hidden sector via kinetic mixing with the SM gauge boson.
The cylindrical configuration, which consists of a circular cylinder as an emitter and a hollow cylinder as a detector, remarkably enhances the geometric factor for an isotropically propagating HP field.
The performance dependency on the aspect ratio of the cavity was examined for various resonant modes.
Exploiting higher-order modes was found to be beneficial when searching for high-mass HPs using a detector designed for the EM solution with the lowest-order field variation.
In addition, the experimental setup does not requires an external magnetic field, which allows for a direct application of the SRF technology to improve the cavity quality factor.
The optimal operating temperature is determined by the given cooling capacity, which limits the RF power injected into the emitter.
By putting all these features together, the experimental sensitivity to the mixing parameter can be improved by multiple orders of magnitude in the $\mu$eV mass range.

\section*{Acknowledgement}
This work was supported by the Institute for Basic Science (IBS-R017-D1-2020-a00/IBS-R017-Y1-2020-a00).


\end{document}